\documentclass[9pt,twocolumn,twoside]{osajnl}

\journal{ol} 

\setboolean{shortarticle}{true}

\usepackage{soul}
\usepackage{color}

\title{Amplified noise nonstationarity in a mode-locked laser based on nonlinear polarization rotation}

\author[1]{Carlos Andres Perilla Rozo}
\author[2]{Philippe Guay}
\author[3]{Jean-Daniel Deschênes}
\author[2,*]{J\'{e}r\^{o}me Genest}

\affil[1]{Universidad Nacional de Colombia, Bogota, Colombia}
\affil[2]{Centre d'optique, photonique et laser, Universit\'{e} Laval, Qu\'{e}bec, Qu\'{e}bec G1V 0A6, Canada}
\affil[3]{Octosig, Qu\'{e}ec, Canada}

\affil[*]{Corresponding author: jgenest@gel.ulaval.ca}

\begin{abstract}
Beat note measurements between a mode-locked and a continuous-wave laser as well as between two mode-locked sources were used to demonstrate that the sub-threshold, cavity filtered, amplified spontaneous emission is not stationary even when a fast mode-locking mechanism, such as nonlinear polarization rotation, is used to generate short pulses. A relatively small gain modulation of a few percents created by high intensity pulses  can produce a significant modulation of the amplified noise once synchronously accumulated over several cavity round-trips, even if the repetition rate is faster than the gain dynamics.
\end{abstract}

\setboolean{displaycopyright}{true}

\begin{document}

\maketitle

The field of mode-locked laser sources also includes a sub-threshold contribution arising from spontaneous emission that is amplified by the gain medium and filtered by the cavity. It was recently shown that this field is temporally distinct from the pulse circulating in the cavity and that it appears in the optical spectrum as shifted from the lasing modes. Since the sub-threshold field experiences the linear cavity while the pulse obviously travel in the saturated nonlinear cavity, this can be used to estimate the cavity's nonlinear phase shift~\cite{PerillaRozo:19}. 

The question of this noisy field's temporal properties then arises. For lasers using a mode-locking scheme which is slow compared to the pulse duration, such as a saturable absorber, it has been shown that the recovery time of the material opens a temporally wide gain window that amplifies spontaneous emission in a nonstationary manner leading to wake modes that can potentially become unstable~\cite{Wang:17}.

For lasers using a fast mode-locking mechanism, stability of this amplified noise is less an issue, but this contribution is observed in dual-comb interference signals and can degrade the signal-to-noise ratio in dual-comb spectroscopy experiments or put additional constraints on the combs parameters such as to make sure the dual-comb beat note is separated from this noisy contribution in the electrical spectrum \cite{DES15}. 

Understanding the temporal and statistical characteristics of this sub-threshold amplified spontaneous emission is therefore important as it could help optimize measurements involving the heterodyning of one or two mode-locked sources.

In this Letter, we used the beat note between an Erbium-doped mode-locked fiber laser and continuous-wave (CW) laser as well as between two similar mode-locked sources to demonstrate that the sub-threshold, cavity filtered, amplified spontaneous emission in a laser employing nonlinear polarization rotation (NPR) is nonstationary. The sub-threshold field is highly modulated even when the pulse rate is faster than the laser dynamics fixed by the relaxation frequency. This is because the relatively small gain modulation imposed by the short pulses is synchronous with the cavity round trip and is thus allowed to accumulate on the amplified noise over its average photon lifetime in the cavity. 

The methods presented here are quite general and could be applied to a wide variety of mode-locked lasers in order to further the characterization of their noise properties.

\begin{figure}[htbp]
\centering
\includegraphics[width=\linewidth]{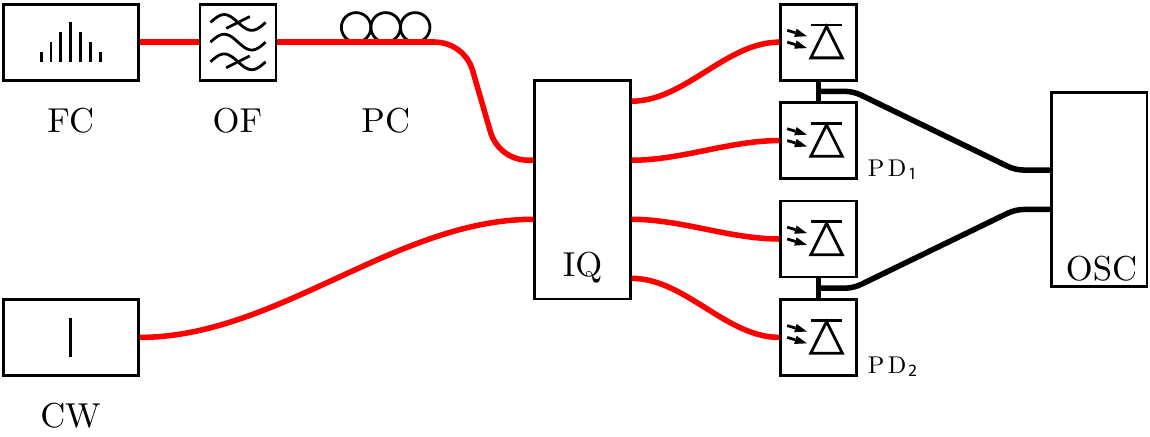}
\caption{Setup for measuring the comb-CW beatnote. FC: Frequency comb. CW: Continuous-wave laser. OF: Optical filter. PC: Polarization controller. IQ: Quadrature demodulation hybrid. PD: Photodetector. OSC: Oscilloscope. }
\label{fig:CW_COMB_Setup}
\end{figure}

\begin{figure}[htbp]
\centering
\includegraphics[width=\linewidth]{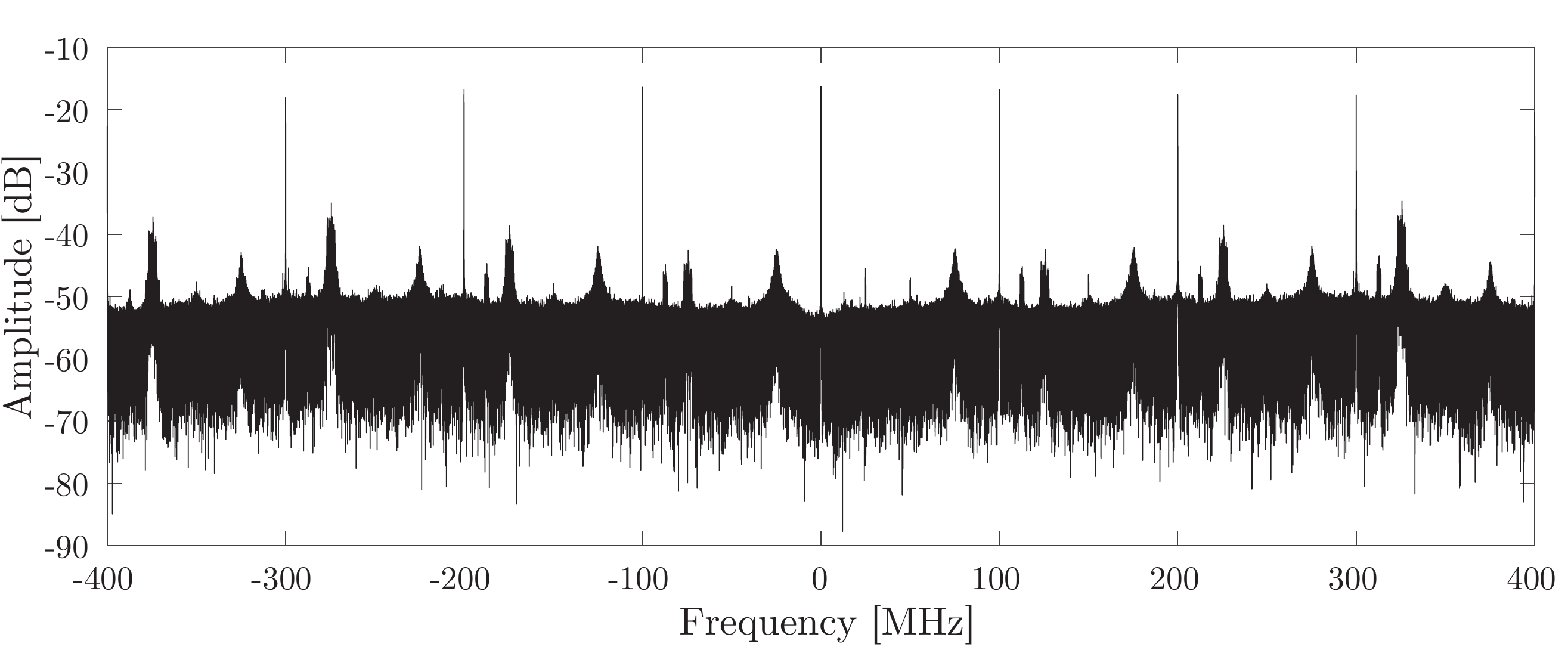}
\caption{Spectrum of the comb-CW beatnote when the CW wavelength is set to $1554.5$~nm.} 

\label{fig:CW_COMB_Beat}
\end{figure}

The beat note between the mode-locked laser (c-Menlo) and a CW laser (Aligent 8164) was first measured using the setup shown in Fig.~\ref{fig:CW_COMB_Setup}. The spectrum obtained from this measurement is given in Fig.~\ref{fig:CW_COMB_Beat}, where the short pulse source field around $1554.5$~nm is filtered (JDS Uniphase TB9) and down-mixed to baseband electrical frequencies by the CW laser. The signal was acquired with an oscilloscope (TeleDyne-LeCroy waveRunner 640Zi). 

In addition to the seven coherent modes observed within the $\pm 350$~MHz detector bandwidth (Thorlabs PDB130C), several features are seen in Fig.~\ref{fig:CW_COMB_Beat}. Details beyond the explanations provided below can be found in~\cite{PerillaRozo:19}.

First, negative radio frequencies are distinguishable from positive ones because an in-phase and quadrature (IQ) detection scheme (Optoplex HB-T0AFAS001-R1) was used~\cite{Bergeron:16}. Some artifacts (at $-180, -80, 120$~MHz,...) are nevertheless left by the imperfect IQ detection leading to only partial cancellation of the negative spectral alias. These will not impact results presented here as they were filtered out in our signal processing steps. 

Of interest for this work are seven "humps" arising from the sub-threshold amplified spontaneous emission filtered by the cavity (at $-120, -20, 80$~MHz,...). These are frequency shifted (by $\approx 20$~MHz) from the lasing modes because the mode-locked part of the field travels in the nonlinear cavity while the sub-threshold noise experiences the linear cavity. This frequency offset is thus a consequence of the cavity's nonlinear phase shift and can be used to characterize it as was done in \cite{PerillaRozo:19}.

Finally, it shall be noted that the spectrum obtained in Fig.~\ref{fig:CW_COMB_Beat} was phase-corrected: the phase evolution of one coherent mode is removed from the entire signal, as in \cite{Deschenes:14}. This not only removes the phase noise from all coherent electrical modes, but also brings them to an harmonic $100$~MHz grid. 

\begin{figure}[t]
\centering
\includegraphics[width=\linewidth]{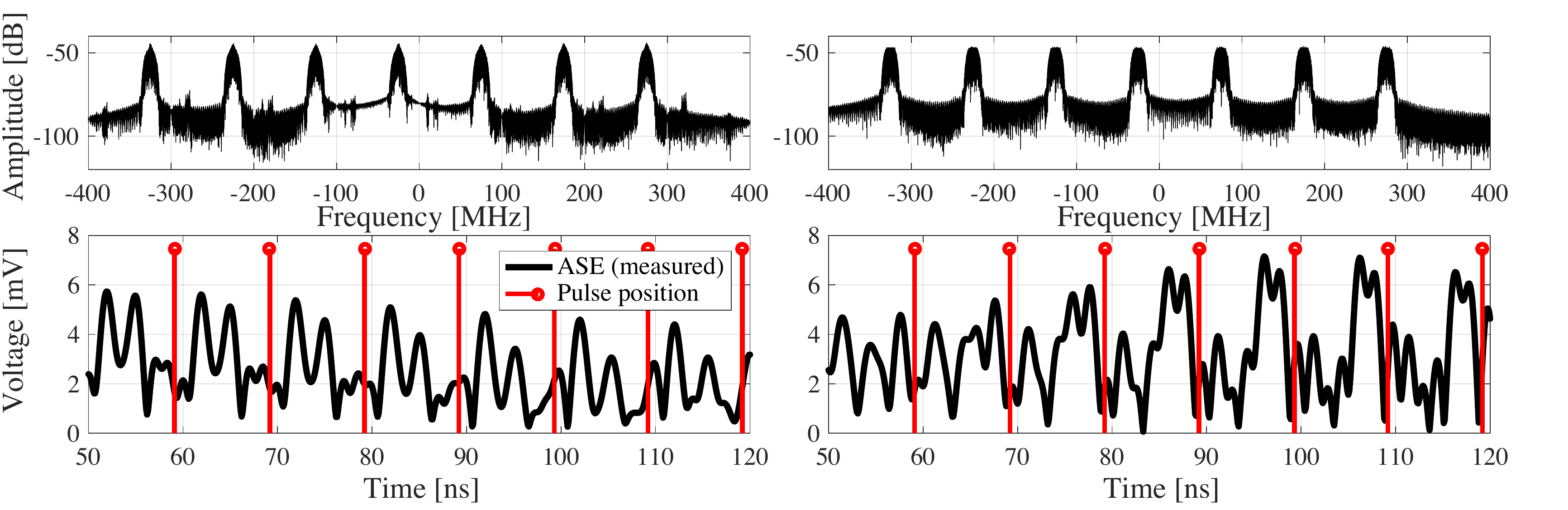}
\caption{Optical noise signal reconstituted by filtering the seven spectral humps (left). Top: Filtered spectrum, bottom: reconstituted time signal's absolute value. Right: Stationary white noise passed through the same filter for comparison purposes. No significant differences are found hence no conclusion can be drawn yet on noise stationarity.}
\label{fig:CW_COMB_Sig}
\end{figure}

In order to make an observation on the optical noise stationarity, one has to isolate the sub-threshold amplified spontaneous emission signal in the measurement. To that effect, a filter was designed to isolate the seven humps and suppress as much as possible the coherent lines. The resulting signals are shown in Fig.~\ref{fig:CW_COMB_Sig} (left) where the spectral (top) and time (bottom) domain representations of the down shifted and filtered sub-threshold field are given. At first glance, it seems that some noisy but periodic structure is visible in the time domain representation, perhaps indicating nonstationarity.

To validate this observation, the same filter was applied to a simulated stationary white noise signal. The right part of Fig.~\ref{fig:CW_COMB_Sig} shows that results are very similar from the optical noise measurements. In fact, on both the measured and simulated time-domain signals, one can see that  some structure is apparently sliding with respect to the signal pulses. This can be explained because the spectral periodicity of the humps, and hence of the filter to isolate them, is different from the tooth signal periodicity. From this first analysis, it appears that no conclusion can be drawn regarding the noise stationarity.

\begin{figure}[t]
\centering
\includegraphics[width=\linewidth]{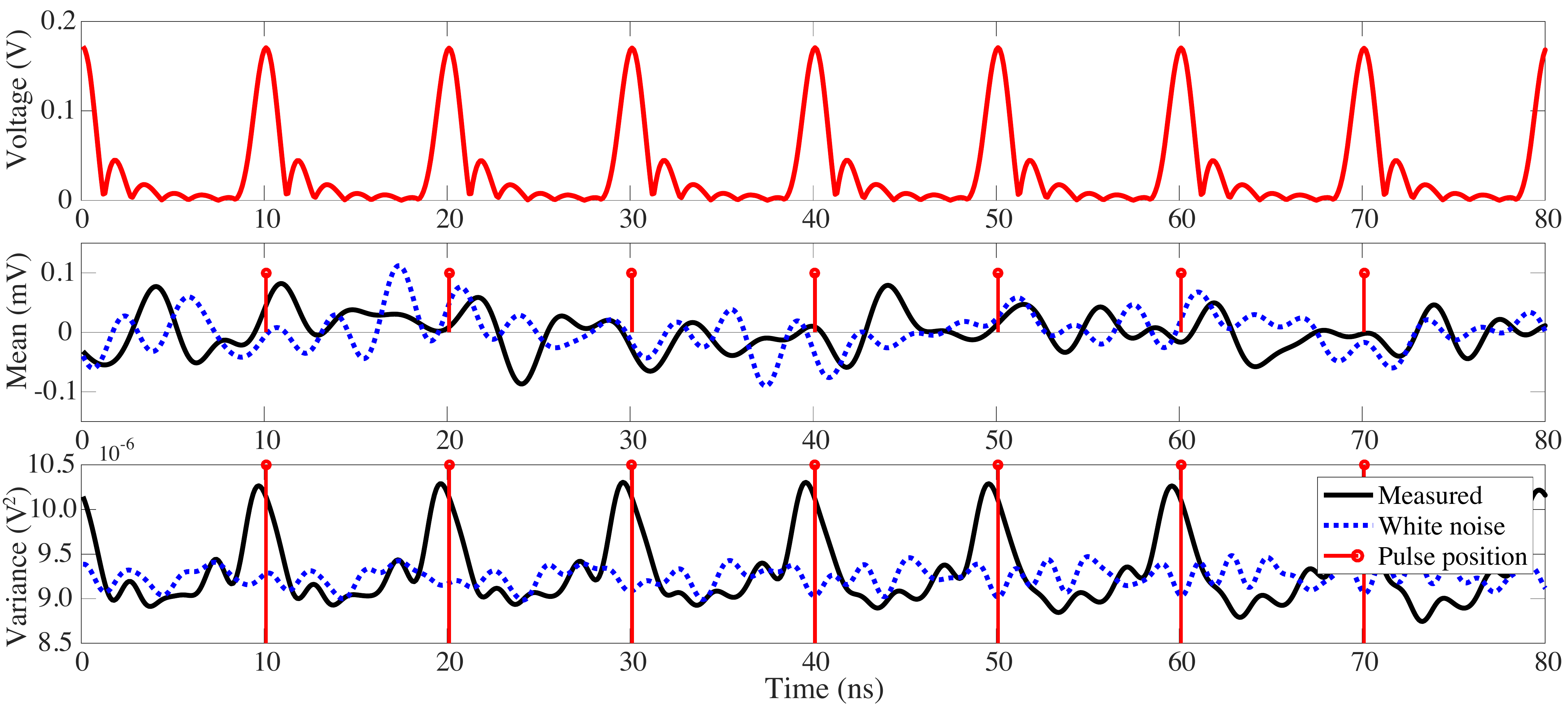}
\caption{Signal statistics over $4000$ slices.  Top, the pulse
train. Middle, average of the filtered amplified noise signal compared to similarly filtered and averaged white stationary noise. Second moment for the filtered amplified noise showing evidence of nonstationarity while the filtered white noise doesn't. }
\label{fig:CW_COMB_Stats}
\end{figure}

Our interpretation of what is happening in the cavity leads to believe that the optical noise nonstationarity should be synchronized with the signal pulses. Indeed, the pulse periodically depletes the gain and the population inversion rebuilds after each pulse passage in the gain medium. For a mode-locked source having a $100$~MHz repetition rate and a measured $800$~kHz relaxation frequency, the depth of this gain modulation is expected to be small, on the order of a few percents, but this might appear in the optical noise temporal profile.

Hence, it makes sense to cut long acquisitions in slices aligned with an integer number of signal pulses and to do statistics across those slices. Fig.~\ref{fig:CW_COMB_Stats} shows the results of computing the average (middle) and the variance (bottom) of $4000$ noise slices. Again, the same process was applied to filtered stationary white noise to ensure observations are significant. The signal pulse train is also shown in Fig.~\ref{fig:CW_COMB_Stats} (top) to highlight the pulses locations.

On the noise average, there is no significant temporal variation when compared to stationary white noise undergoing the same processing. This is expected since the sub-threshold amplified spontaneous emission is a zero mean optical field. However, for the second moment which is related to noise power, there is an increasing trend just before the occurrence of a signal pulse. Filtered white stationary noise does not show these characteristics, hinting towards the fact that ASE nonstationarity is indeed measured.

Although the observation is somewhat consistent with our understanding of the cavity physics: population inversion and thus amplified noise builds before a pulse and decreases when the pulse depletes the gain, the measurement is not extremely convincing. First, the temporal resolution is limited by the detector response time which is $1.6$~ns in this case. Second, the nonstationarity may be induced by an electrical phenomenon. For instance, the signal pulse might induce nonlinearity in the detection chain and the noise signal would see this gain saturation. With this setup, we cannot tell for sure if the nonstationarity is on the optical signal or if it appears in the electrical conversion. 

Even with the somewhat limited temporal resolution, it can be noted that there is delay of $\approx 500$~ps between the noise peak value and the location of the pulses. 

To make sure the observed nonstationnarity is on the noisy optical signal, a second measurement making use of linear optical sampling \cite{Dorrer2003LinearOS,Coddington:09} was made.  This was achieved by replacing the CW laser in Fig. \ref{fig:CW_COMB_Setup} by a second mode-locked laser similar to the first but slightly detuned in repetition rate. The setup became thus a dual-comb interferometer, used in a manner very similar to dual-comb spectroscopy \cite{Coddington:16}. For this experiment, we however aimed to sample the sub-threshold amplified noise of one comb with the pulses of the other, which is acting as a local oscillator. The measurement was no longer limited by detector response time because of this optical equivalent time sampling.

Instead of performing a Fourier transform of a full interferogram, as is customary the case in dual-comb interferometry, we here took Fourier transforms on short interferogram sections. Having several interferograms cut into small sections, we can do statistics across each section.

\begin{figure}[b]
\centering
\includegraphics[width=\linewidth]{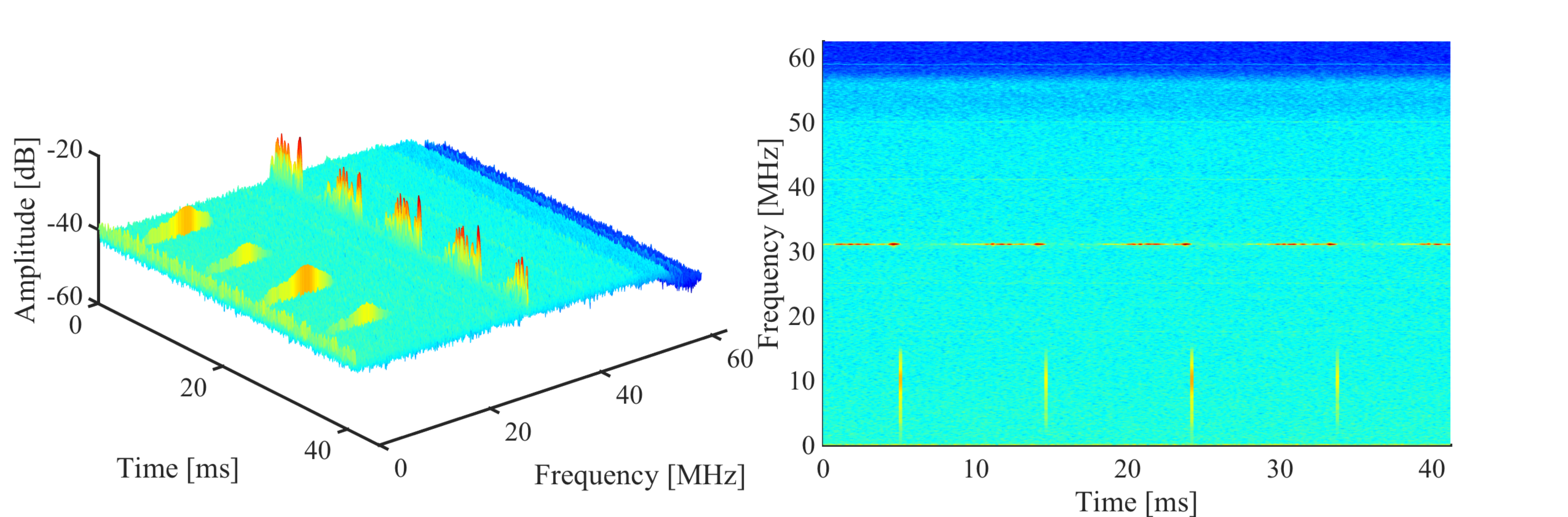}
\caption{Spectrogram for a dual-comb beat. Classical wideband dual-comb signal (pulse to pulse interference) around $10$~MHz and pulse to optical noise beat note around $31$~MHz. Resolution bandwidth RBW$=48$~kHz.}
\label{fig:DUAL_COMB_3D}
\end{figure}

Fig.~\ref{fig:DUAL_COMB_3D} shows a spectrogram of the second order moment when the repetition rate difference is $\Delta f_r = 104$~Hz. Here, the short-time Fourier transform length and hence the resolution bandwidth was adjusted so that the pulse and noise feature amplitudes were comparable to ease visualization. $25$ interferograms and a resolution bandwidth of 48 kHz were used to compute the second order moment. 

On Fig. \ref{fig:DUAL_COMB_3D}, the wideband comb-comb beatnote can be seen between $5$~MHz and $15$~MHz. This signal is localized in the time domain, indicating that the local oscillator is sampling pulses.

The beat note arising from the sampling of one comb's noise with the pulse of the other is seen around $31$~MHz, showing clear evidence of nonstationarity. Once again the $\approx21$~MHz shift between the pulse-pulse and pulse-noise beat notes is indicative of the cavity's nonlinear phase shift, allowing to confirm the nature of this signal observed at $31$~MHz. 

The fact that the optical noise spectral feature is spectrally narrow around 31 MHz can at first be seen as problematic. In theory, the cavity filtered amplified spontaneous emission should have a bandwidth comparable to lasing modes. In \cite{PerillaRozo:19}, the amplitude and shift relative to the closest comb mode of each hump was characterized as a function of wavelength over the full optical bandwidth of the source. Using this data, it was possible to simulate the expected width of the comb to sub-threshold amplified spontaneous emission beatnote, and it is perfectly consistent with what is observed in the spectrogram, that is a full width half maximum around 1.5~MHz. The simulation showed that this is mainly due to the high spectral selectivity on the amplitude for the sub-threshold amplified spontaneous emission, such that only a narrow spectral range contributes to what is seen on the spectrogram of Fig. \ref{fig:DUAL_COMB_3D}.

\begin{figure}[t]
\centering
\includegraphics[width=\linewidth]{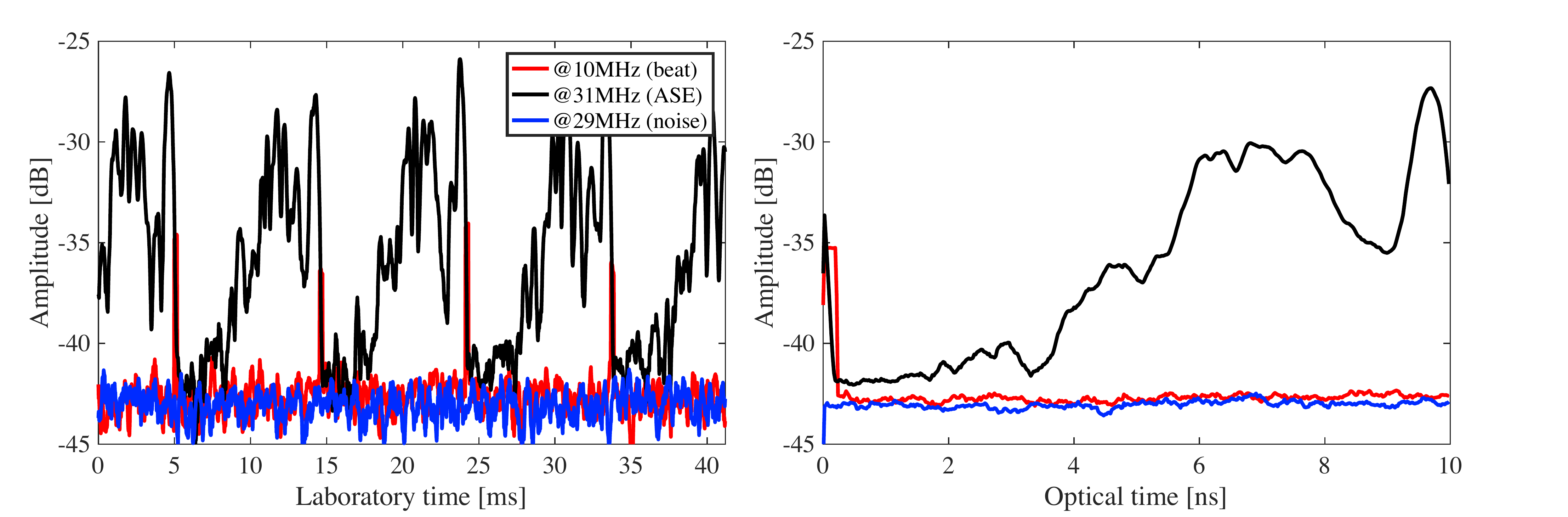}
\caption{Spectrogram slices at $10, 31 \text{and} 29$~MHz showing the temporal evolution of the pulse (red), the amplified spontaneous emission (black) and the background measurement noise respectively (orange).}
\label{fig:DUAL_COMB_2D}
\end{figure}

To better observe the temporal evolution of the signals in the spectrogram of Fig. \ref{fig:DUAL_COMB_3D}, slices were taken at $10$, $31$ and $29$~MHz to visualise separately the dual-comb beat note, the sampled amplified spontaneous emission and the background measurement noise, respectively. These are shown in blue, red and green in Fig. \ref{fig:DUAL_COMB_2D}. The left panel shows four successive periods while the right panel shows a single period, but with $25$ averages to minimize additive noise. 

The sub-threshold amplified spontaneous emission nonstationnarity is quite obvious and showing a similar trend to what was obtained with the mode-locked to CW beat note. The optical noise signal is strongly reduced after a pulse and rebuilds progressively until the next one. Thanks to the higher temporal resolution provided by linear sampling, the delay between the  amplified emission peak and the pulse arrival was now estimated to $330$~ps, which is also consistent with the first measurement presented. Together, the two measurements confirm the optical nature of the phenomenon and clearly demonstrate that sub-threshold cavity filtered amplified spontaneous emissions is not stationary in a mode-locked laser based on NPR. 

\begin{figure}[h]
\centering
\includegraphics[width=\linewidth]{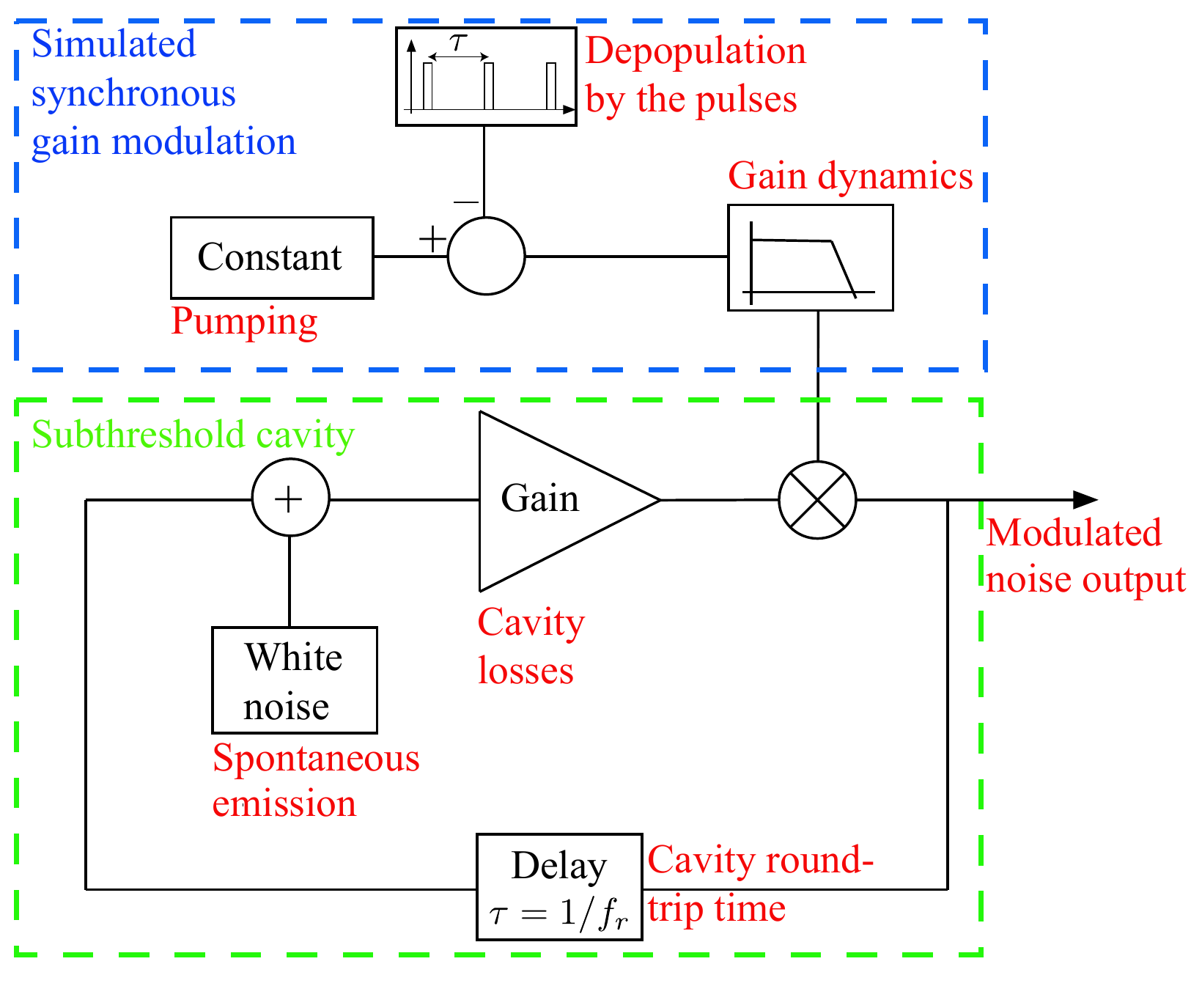}
\caption{Block diagram of the laser cavity modelisation.}
\label{fig:BLOCK_DIAGRAM_model}
\end{figure}

One stricking feature of Fig. \ref{fig:DUAL_COMB_2D} is the fact that the observed nonstationarity has a modulation depth larger than $14$~dB, at a repetition rate of $100$~MHz. This seems to be in sharp contrast with the idea that rare-earth doped fiber lasers have a slow dynamic such that the gain should not follow at this speed.

The observations were validated using a MATLAB Simulink modelisation of the laser cavity. The simulation was not intended to fully capture the laser dynamics. It should rather be interpreted as a simple model to coarsely validate our understanding of the observed phenomenon. A block diagram of the model is shown in Fig. \ref{fig:BLOCK_DIAGRAM_model}.  The sub-threshold amplified spontaneous emission is modeled as band-limited white noise that is generated in a feedback loop. The signal experiences the gain and losses of the cavity and a delay corresponding to the round-trip time $(\tau = 1/f_r)$. The cavity gain is multiplied by the output of a first-order filter that roughly simulates the gain medium modulation dynamic. The cut-off frequency of this filter is set to the laser's relaxation frequency which was measured at 800~kHz. The gain medium is pumped with a constant signal from which is subtracted a pulsed signal simulating the depopulation of the upper level by the pulses in the cavity. The frequency of theses pulses if defined by the repetition rate of the laser at $f_r=1/\tau$.

\begin{figure}[h]
\centering
\includegraphics[width=\linewidth]{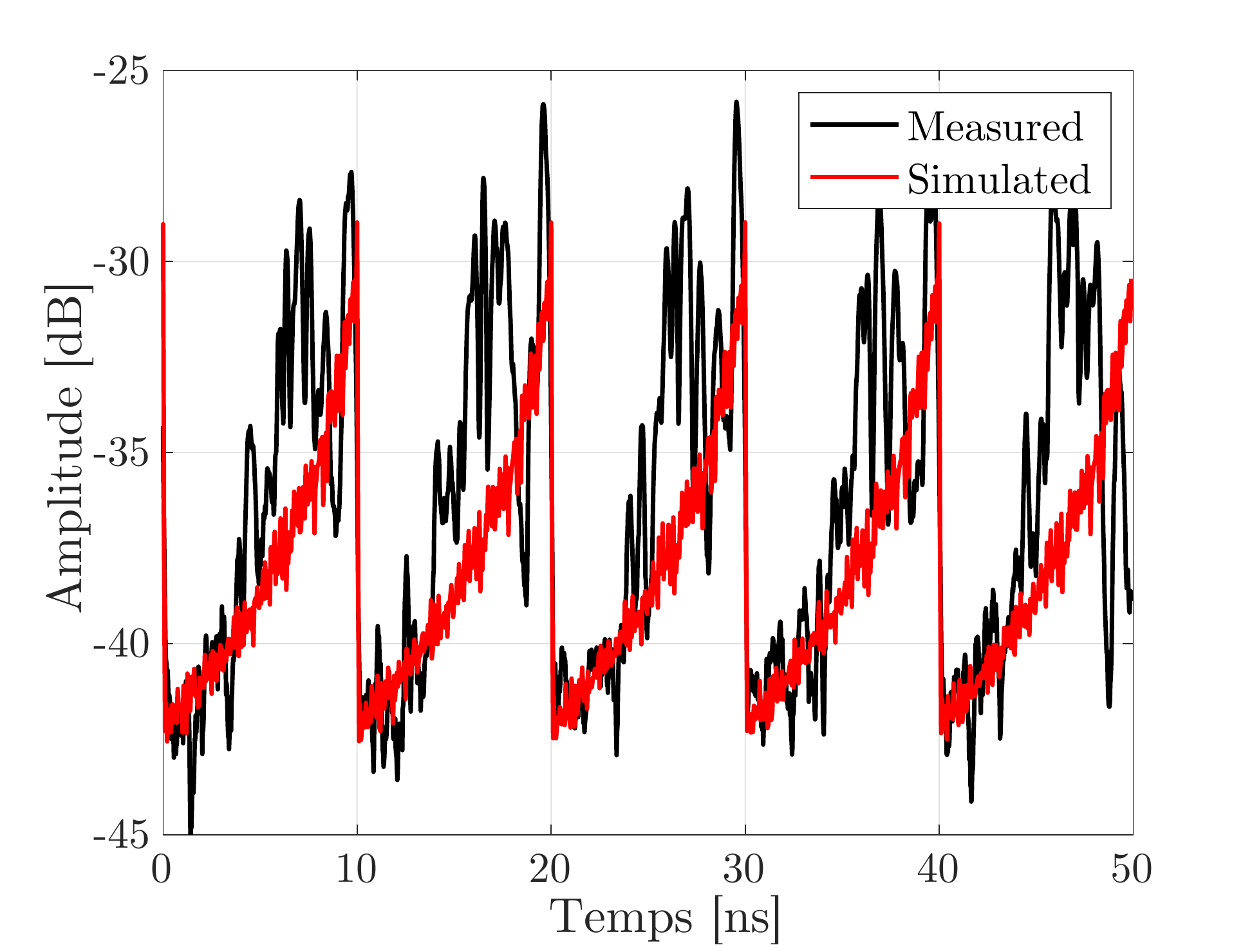}
\caption{Spectrogram slice at 31 MHz showing the temporal evolution of the amplified spontaneous emission for the experimental (black) and simulated data (red).}
\label{fig:DUAL_COMB_2D_sim}
\end{figure}

The parameters of the model (losses, gain modulation depth) were adjusted to match the experimental data. First, the losses in the cavity, which determines its finesse, were set to 9\% to agree with the 1.5 MHz spectral width of the sub-threshold amplified spontaneous emission observed in Fig. \ref{fig:CW_COMB_Beat}. Second, the gain modulation due to stimulated emission in the cavity was set to 8\% to match the experimentally observed modulation depth of around 14 dB.

Not all features are captured, but reasonable correspondence in the general gain modulation behavior is observed between the experimental and simulated data, as can be seen in Fig. \ref{fig:DUAL_COMB_2D_sim}. This confirms that significant modulation of sub-threshold noise faster than the gain dynamics is possible in a mode-locked laser and that this comes from the accumulation of a synchronous gain modulation overall several cavity round trips ($\approx 15$ in the present case).








In conclusion, nonstationarity of the sub-threshold spontaneous amplified emission was observed with direct comb and dual-comb approaches. It was shown that a small gain modulation created by high intensity pulses synchronously accumulated over several cavity round-trips produce a 95\% modulation of the amplified noise. The results, validated with a simple modelisation of the laser cavity, show that the amplified noise can be modulated at the repetition rate of the laser, even though the gain dynamics is much slower.

\section*{Funding Information}

This work was supported by the Natural Sciences and Engineering Research Council of Canada (NSERC) and by Universidad Nacional de Colombia.

\bibliography{sample}


\end{document}